# Possible phase separated bulk superconductivity in $Ru_{0.9}Sr_2YCu_{2.1}O_{7.9}$

V.P.S. Awana, T. Kawashima and E. Takayama-Muromachi*

Superconducting Materials Center (Namiki Site), National Institute for Materials Science, 1-1 Namiki, Tsukuba, Ibaraki 305-0044, Japan

The $Ru_{0.9}Sr_2YCu_{2.1}O_{7.9}$ sample being synthesized by high-pressure high-temperature solid-state reaction underwent (weak) ferromagnetic transition at ~150 K followed by superconducting transition at ~30 K. It showed clear Meissner signal in the field-cooled process up to external magnetic field of a few hundred Oe (~300 Oe). The magnetic susceptibility data and magnetic hysteresis data could be explained assuming that the sample was a macroscopic mixture of a superconductor and a (weak) ferromagnet. Any anomalous behaviour implying the coexistence of superconductivity and magnetism on a microscopic scale was not observed.

PACS No. 74.25Fy, 74.25Ha, 74.72 Jt,

Recent discovery of the "superconducting ferromagnet" of $LnSr_2GdCu_2O_8$ (Ln/Ru-1212, Ln = Gd, Eu, Y) is of tremendous interest [1]. In these compounds, ordering of Ru spins occurs at $T_N$ = 120~150 K followed by the superconducting transition at $T_c$ = 10~40 K. A similar phenomenon was also observed in $RuSr_2Gd_{1.5}Ce_{0.5}Cu_2O_{10}$ (Gd/Ru-1222) [2]. Bulk nature of the ferromagnetic order parameter in Gd/Ru-1212 was evidenced from muon-spin-resonance (µSR) and electron-spin-resonance (ESR) experiments [1,3]. However, there are still serious discrepancies among magnetic properties obtained by different experimental tools. Neutron diffraction experiments for Ln/Ru-1212 (Ln = Gd, Eu, Y) indicated that they have a G type antiferromagnetic structure with $\mu_{Ru}$ ~ $1\mu_B$ and ferromagnetism in these phases is due to canting of the Ru moments [4-7]. Recent DC magnetic measurements for Gd/Ru- and Eu/Ru-1212 have cast doubt for the model proposed by the neutron studies from various points of view [8].

Bulk nature of superconductivity was initially criticised due to lack of Meissner signal in Gd/Ru-1212 and rather a crypto superconducting phase was proposed [9,10]. Recently, however, Bernhard et al. reported a sizable Meissner signal in a field-cooled (fc) process for a Gd/Ru-1212 sample as an evidence of a bulk Meissner phase [11]. According to them, the bulk Meissner phase develops below a certain temperature, $T^{ms}$ which is substantially lower than the superconducting transition temperature, $T_c$, and a spontaneous vortex phase exists in the intermediate temperature region, $T^{ms} < T < T_c$. Quite similar picture on the coexistence of superconductivity and magnetism has been proposed for Gd/Ru-1222 phase [12].

As far as the phase purity is concerned, good amount of work is done on Gd/Ru-1212 including synchrotron x-ray [13], neutron powder [6] diffractions, and high resolution transmission electron microscopy [13]. Yet the question of phase purity is not resolved to the satisfactory level [14]. Not only various phase pure non-superconducting samples do exist [15], but also the reproducibility of superconducting compounds with same heat treatments is reported in doubt [9]. Some of these puzzles lie with the fact that solid solutions of $Ru_{1-x}Cu_x$-1212, which can be superconducting with x > 0.5, but not necessarily magnetic may precipitate within the stoichiometric Ru-1212 composition [16]. Because Ru and Cu have close scattering cross-sections for neutron, hence neutron diffraction study is not very helpful for such a problem.

It seems that we still need fundamental data, in particular on magnetic properties, for the final conclusion on the coexistence of superconductivity and magnetism. To see the magnetic behaviour, Gd/Ru-1212 is not a proper system because of the presence of magnetic Gd ions ($8\mu_B$) which hinder in knowing the exact magnetic contributions form the Ru ions and form superconductivity. Ru-1212 can be formed for non-magnetic Y instead of Gd, but only with high-pressure high-temperature (HPHT) synthesis technique [5,17,18]. It is our aim here to study the magnetic properties of the superconducting Y/Ru-1212 sample which was prepared by the HPHT method. We observed Meissner signal in the fc process until few hundred Oe external fields for this sample. This result presents a striking contrast to Gd/Ru-1212 for which fc diamagnetic signal was seen only till few Oe fields or even no signal.

Sample of composition $Ru_{0.9}Sr_2YCu_{2.1}O_{7.9}$ was synthesised through a HPHT solid-state reaction with ingredients of $Y_2O_3$, $SrO_2$, $SrCuO_2$, $RuO_2$ and $CuO$. Details of sample synthesis are given elsewhere [17,18]. Slightly Ru-poor starting composition was selected because a single-phase sample is obtained from this composition without the contamination of $SrRuO_3$ [17]. X-ray powder diffraction patterns were obtained by a diffractometer (Philips-PW1800) with Cu $K_\alpha$ radiation. DC susceptibility data were collected by a SQUID magnetometer (Quantum Design, MPMS).

$Ru_{0.9}Sr_2YCu_{2.1}O_{7.9}$ crystallised in a single-phase form in space group *P4/mmm* with lattice parameters $a = b = 3.816(1)$ Å, and $c = 11.514(3)$ Å [18]. Figure 1 shows both zero-field-cooled (zfc) and fc magnetic susceptibility versus temperature ($\chi$ vs. *T*) plots for the $Ru_{0.9}Sr_2YCu_{2.1}O_{7.9}$ sample, in various external fields, H, of 50, 70, 100, 300, and 1000 Oe. As seen from this figure the zfc and fc magnetization curves show a rapid increase near 150 K followed by a significant branching at around 145 K. The branching is indicative of the long range magnetic order of the Ru moments. Neutron diffraction studies revealed that the Ru moments order antiferromagnetically at $T_N = 133$ K and 120 K, in Gd/Ru-1212 and Eu/Ru-1212, respectively [4,6,7]. In a recent neutron study on Y/Ru-1212 prepared by the HPHT process, similar magnetic order was observed at $T_N = 149$ K [5], which is in close agreement to the current value. With an increase in applied field (10 Oe < H < 1kOe) basically no change is observed in the magnetic transition temperature.

The zfc part of magnetic susceptibility at low T below 30 K, shows clear diamagnetism up to H ~ 300 Oe. The diamagnetism is field dependent, and almost disappears at H = 1000 Oe. The diamagnetic signal onset temperature is described as superconducting transition temperature ($T_c$) although the transition is rather broad. Worth noting is the fact that the diamagnetic signal observed in the zfc process does not saturate down to 5 K.

The fc part of the magnetic susceptibility remains positive down to 5 K. However, Meissner signal is observed clearly as a dip below ~ 30 K. The dip in the fc susceptibility is dependent on the external field, higher is the field and less is the dip. For applied magnetic field of 1000 Oe, the dip disappears almost completely. The dip in the fc curve corresponds well to the decrease of the susceptibility in the zfc curve with almost the same onset temperature. Though a negative susceptibility is not observed in the fc process, the observation of the clear dip guarantees the bulk nature of superconductivity in the sample. We roughly evaluate the dip by defining $\Delta\chi$ as a difference of susceptibility between $T_c$ (onset) and 5 K, which is plotted in the inset of Fig. 1. It is clear that $\Delta\chi$ decreases with an increase in H. However, $\Delta\chi$ has a nonzero value even at H = 300 Oe. Estimated Meissner superconducting volume fraction (calculated from the $\Delta\chi$ value) is nearly 15 % at H = 50 Oe and above 4 % at H = 300 Oe. The fc susceptibility is almost saturated below about 10 K in contrast to the non-saturated zfc one resulting in increase of the difference between the fc and the zfc values with decreasing temperature. This probably reflects the effect of pinning of vortexes, i.e., we have to suppose fairly strong pinning for the present system.

Bernhard et al. also reported a dip in the fc process for a Gd/Ru-1212 sample [11]. However, clear dip was observed only at very low applied fields (< 10 Oe). According to their scenario on the coexistence of superconductivity and ferromagnetism, a spontaneous vortex phase (SVP) is formed when the spontaneous magnetization, $4\pi M$, exceeds the lower critical field $H_{c1}$ (i.e., $4\pi M > H_{c1}$). On the other hand, the Meissner phase will become stable if $4\pi M < H_{c1}$. Since $H_{c1}$ depends on temperature with being zero at $T_c$ while $4\pi M$ is practically constant for temperatures near or blow $T_c$, the Meissner state will develop below a certain temperature, $T^{ms}$ which is substantially lower than $T_c$, and a SVP exists in the intermediate temperature region, $T^{ms} < T < T_c$. When the external magnetic field is applied, the Meissner phase will occupy narrower area with $4\pi M + H < H_{c1}$. For the Gd/Ru-1212 sample, Bernhard et al. estimated $4\pi M$ of the order of 50-70 Oe and $H_{c1}(T=0)$ of the order of 80-120 Oe. The difference between the two values (30-50 Oe) is not so large and this was claimed to be a reason for the Meissner signal disappearing under a higher external magnetic field, $H_t > 35$ Oe. Similar picture has been proposed for Gd/Ru-1222 phase [12].

If the scenario mentioned above is simply applied to the present system, we have to assume quite high $H_{c1}$ of the order of 350 Oe. Moreover, it should be stressed that if we subtract the ferromagnetic contribution in Figure 1, the remaining magnetic susceptibility curve looks very normal compared with those of high $T_c$ oxides. Although superconducting transition is rather broad in the present sample, such a broadening often occurs in an inhomogeneous system. In other words, the susceptibility data in Figure 1 can be explained assuming a macroscopic mixture of a superconductor and a ferromagnet. It is worth discussing here that the DC electric resistivity of the present Y/Ru-1212 sample is quite high and zero resistivity is not usually attained even when the sample showed a large diamagnetism in a low temperature region. The present sample also showed semiconducting behaviour (~2 $\Omega$/cm at 300K and ~50 $\Omega$/cm at 10 K) with small downturn at 8 K [18]. This fact supports the macroscopic mixing state of the sample. Although zero resistivity was attained after high-oxygen-pressure post annealing [5,17], it resulted in decomposition of the Y/Ru-1212 phase and superconducting volume fraction is not increased at all after the post annealing [17].

Figure 2 depicts the magnetization loop for the $Ru_{0.9}Sr_2YCu_{2.1}O_{7.9}$ sample at 5 K with the applied fields in the range of 1000 Oe < H < 1000 Oe. At the present stage, it is not known exactly what kind of anomaly is expected in the magnetization curve for the coexistence system of superconductivity and magnetism, i.e., for SVP and the Meissner phase. Sonin and Felner carried out theoretical analysis and proposed an equilibrium magnetization curve expected for the system in question [12]. However, as clear magnetic hysteresis is seen in Figure 2, the present system is far from the equilibrium state (it is also worth noting that fairly strong pinning effect is suggested form the magnetic susceptibility data). The experimental hysteresis loop in Figure 2 looks a mere superimposition of superconducting and ferromagnetic hysteresis without any anomalies, and again, can be explained assuming that the sample consisted of a macroscopic mixture of a superconductor and a ferromagnet. The $H_{c1}$ value estimated from Figure 2 is reasonable with the order of 100 Oe.

In Figure 3 are shown the magnetization curves at various temperatures of 5, 20, 50, 100, 120 and 150 K, in applied fields of $-70$ kOe $<$ H $<$ 70 kOe. According to the neutron diffraction experiments, Ln/Ru-1212 (Ln = Gd, Eu, Y) phases order below $T_N$ in a G type antiferromagnetic structure with $\mu_{Ru} \sim 1 \mu_B$ along the c-axis and canting of the moments gives a small ferromagnetic component less than 0.3 $\mu_B$ [4-7]. From Figure 3, the magnetization at 70 kOe and 5 K is 1.17 $\mu_B$, and the extrapolation of the high magnetic field data at 5 K to H = 0 gives $M_0 \sim 1 \mu_B$. These values are in good agreement with previous magnetization reports for the Ln/Ru-1212 (Ln=Gd, Eu, Y) phases [1,5,7]. $M_0$ is close to $\mu_{Ru}$ proposed by the neutron diffraction experiments. The agreement of the two values means that the Ru moments aligned parallel to the external magnetic field with an external field of ~ 40 kOe. However, this seems somewhat curious if we consider $T_N$ as high as ~150 K. The antiferromagnetic correlation should be of the order of 150 K and it is natural to assume that a very high magnetic field with corresponding strength is needed in order to align the Ru moments completely parallel.

According to a recent report for Gd/Ru-1212 and Eu/Ru-1212, high temperature magnetic susceptibility data gave $\mu_{Ru} \sim 2.5 \mu_B$ for both phases which is 2.5 times larger than the neutron value. Using this value and considering within the canted antiferromagnetism regime, the low-temperature magnetization curves in Figure 3 may be divided into two parts H < 40 kOe and H > 40 kOe. The data for H < 40 kOe mainly reflects the process that the net ferromagnetic moment aligns parallel to the external magnetic field changing its direction from the easy axis of magnetization (or easy plane of magnetization). According to the magnetic structure proposed by the neutron experiments, the easy axis of magnetization for the net ferromagnetic moment is the *a-b* plane and this plane is not always parallel to the external magnetic fields in a polycrystalline ceramic sample. It may be worth noting here that recent ferromagnetic resonance study suggested an extremely large easy plane anisotropy of ~110 kOe [3], though it is much higher than the present value of ~40 kOe.

In the second process for H > 40 kOe, canting angle of the Ru moment may increase exclusively with H resulting in the linear increase of the magnetization. According to this scenario, $M_0$ is not the "saturation" magnetization but the spontaneous magnetization (the same interpretation has been made on Gd/Ru-1222 in Ref. 10), giving internal dipolar magnetic field of ~700 Oe (= $4\pi M_0$). This value of the magnetic field is in good agreement with the result of the μSR measurement for Gd/Ru-1212 [1] but is one order of magnitude larger than that by the neutron diffraction experiments, and in addition, it will require a large canting angle, which may be unusual [6]. At the present stage, it is very difficult to propose a definite model for the magnetism of Ru-1212 because of the serious discrepancies among the magnetic data obtained by different experimental tools. It seems that single-crystal measurements are needed for the final conclusion.

In summary, our Y/Ru-1212 sample prepared under high pressure appeared to undergo (weak) ferromagentic transition at ~150 K followed by bulk superconducting transition at ~30 K. Clear Meissner signal was observed in the fc process up to H = 300 Oe, in contrast to the earlier reports for various Ru-1212 samples which showed Meissner signal only under very low magnetic field or even no signal. The magnetic susceptibility and magnetization data can be explained assuming a macroscopic mixture of a superconductor and a ferromagnet, without any anomalies, which could imply the coexistence of superconductivity and magnetism on a microscopic scale. The ferromagnetic properties obtained by the DC magnetic measurements were not fully consistent with the magnetic structure proposed by neutron diffraction experiments.

**FIGURE CAPTIONS**

Figure 1. Magnetic susceptibility versus temperature ($\chi$ vs. T) plots for $Ru_{0.9}Sr_2YCu_{2.1}O_{7.9}$ sample, in various applied fields of 50, 70, 100, 300, and 1000 Oe, inset shows the difference of magnetic susceptibility ($\Delta\chi$) between $T_c$ (onset) and 5 K for field-cooled (fc) transition in various fields.

Figure 2. M vs. H plot for the $Ru_{0.9}Sr_2YCu_{2.1}O_{7.9}$ compound at 5 K, the applied field H are in the range of $0 = H = 1000$ Oe.

Figure 3. M vs. H plot for the $Ru_{0.9}Sr_2YCu_{2.1}O_{7.9}$ compound at $T = 5, 20, 50, 100, 120$ and 150 K, the applied field H are in the range of $-70$ kOe $< H < 70$ kOe.

*Figure 1, Awana et al. Phys. Rev. B.*

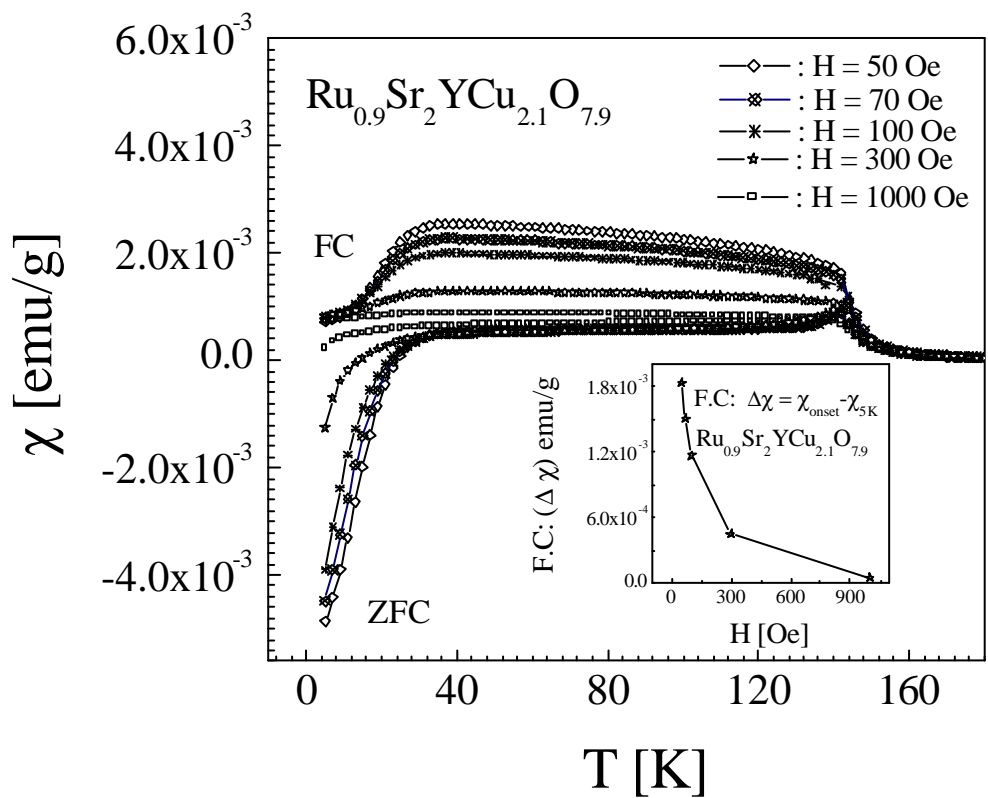

*Figure 2, Awana et al. Phys. Rev. B.*

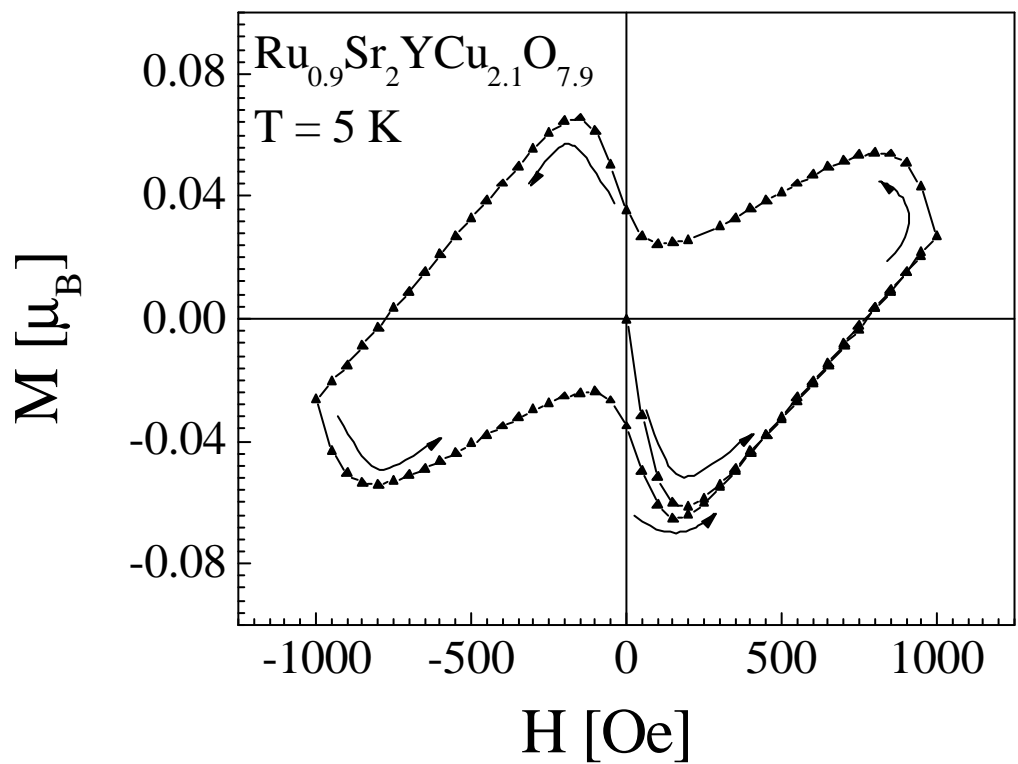

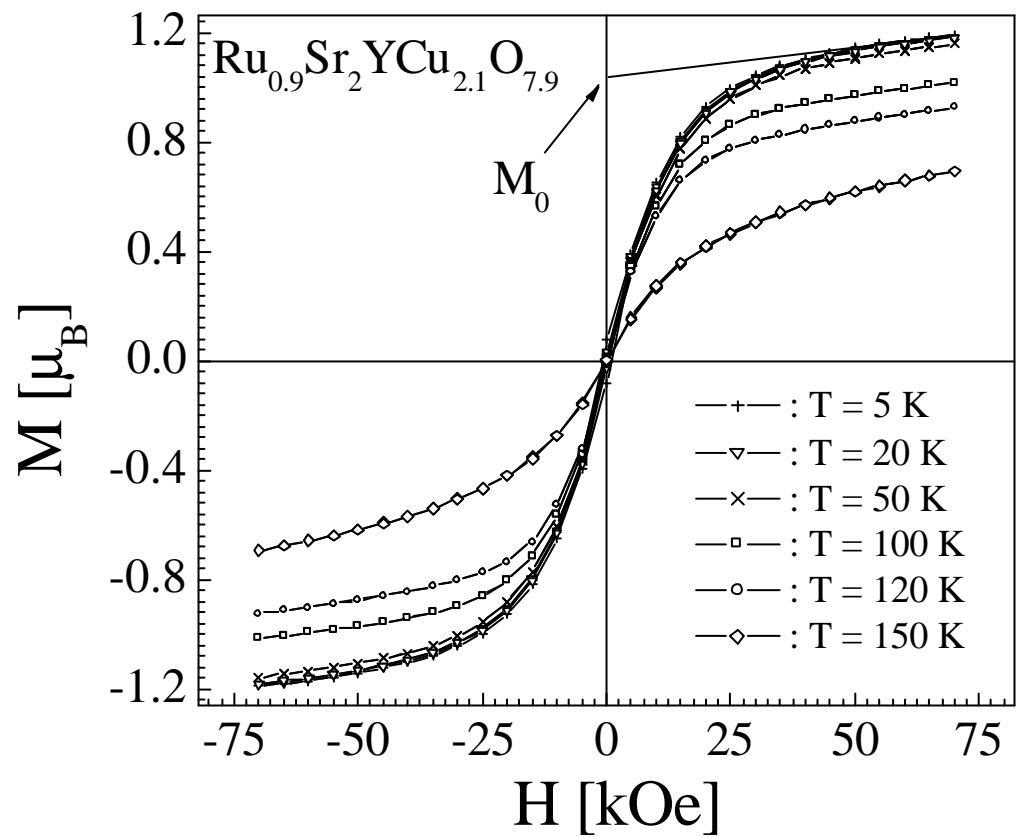